\documentclass[preprint,showpacs,preprintnumbers,amsmath,amssymb]{revtex4}

\usepackage{bm}

\begin{document}


\title{\textbf{Topological regularization and self-duality\\
in four-dimensional  anti-de Sitter gravity}}

\author{Olivera Mi\v{s}kovi\'{c}}
\email{olivera.miskovic@ucv.cl}
 \affiliation{Instituto de F\'{\i}sica, P. Universidad Cat\'{o}lica de
Valpara\'{\i}so,\\ Casilla 4059, Valpara\'{\i}so, Chile}

\author{Rodrigo Olea}
\email{rodrigo_olea_a@yahoo.co.uk}
\affiliation{Instituto de F\'{\i}sica, P. Universidad Cat\'{o}lica de
Valpara\'{\i}so,\\ Casilla 4059, Valpara\'{\i}so, Chile}
\affiliation{Max-Planck-Institut f\"{u}r Gravitationsphysik, Albert-Einstein-Institut\\
Am M\"{u}hlenberg 1, 14476 Golm, Germany}

\date{\today}

\begin{abstract}

It is shown that the addition of a topological invariant
(Gauss-Bonnet term) to the anti-de Sitter (AdS) gravity action in
four dimensions recovers the standard regularization given by
holographic renormalization procedure. This crucial step makes
possible the inclusion of an odd parity invariant (Pontryagin term)
whose coupling is fixed by demanding an asymptotic (anti) self-dual
condition on the Weyl tensor. This argument allows to find the dual
point of the theory where the holographic stress tensor is related
to the boundary Cotton tensor as $T_{j}^{i}=\pm (\ell ^{2}/8\pi G)C_{j}^{i}$,
which has been observed in recent literature in solitonic solutions
and hydrodynamic models.

A general procedure to generate the counterterm series for AdS
gravity in any even dimension from the corresponding Euler term is
also briefly discussed.

\end{abstract}


\maketitle


\section{Introduction}

In the gravity side of the anti-de Sitter/Conformal Field Theory
(AdS/CFT) correspondence, the relevant information that realizes
this duality is encoded in the finite part of the boundary
stress tensor \cite{Maldacena}. That identification requires the cancelation of the
infrared divergences in the bulk theory made by holographic
renormalization procedure \cite{dHSS}, which is based on an asymptotic analysis
of the metric in Fefferman-Graham (FG) coordinate system \cite{Fefferman-Graham}
\begin{equation}
ds^2=\frac{\ell^2}{4\rho^2}\,d\rho^2+\frac{1}{\rho}\,g_{ij}\,dx^{i}dx^{j}, \label{FG}
\end{equation}%
where $h_{ij}=g_{ij}/\rho$ corresponds to the boundary metric.
For asymptotically AdS (AAdS) spaces, $g_{ij}(x,\rho )$ accepts a regular
expansion near the boundary $\rho =0$ , i.e.,
$ g_{ij}(x,\rho )=g_{(0)ij}+\rho g_{(1)ij}+\cdots $. Solving the Einstein equations in this frame leads to
the holographic reconstruction of the spacetime from a given boundary data
$g_{(0)ij}$, what is essential to determine the series of intrinsic counterterms ${\cal L}_{ct}$
which renders finite the boundary stress tensor \cite{Ba-Kr}.

However, the algorithm which produces ${\cal L}_{ct}$ becomes extremely complex as
the spacetime dimension increases, such that there is no a closed formula for counterterms
for an arbitrary dimension. This argument motivates the search for alternative approaches.

On the other hand, any other regularization scheme, even if properly removes the
asymptotic divergences, might spoil the holographic interpretation
of the theory within the AdS/CFT framework because of different boundary conditions.

In particular, a regularization mechanism for AdS gravity in any dimension which consists
in the addition of counterterms that depend on the extrinsic
curvature $K_{ij}$ (Kounterterms method) has been
recently proposed \cite{OleaJHEP, OleaKT}. In this case, the on-shell variation of the
regularized action $I_{reg}$ contains terms of the type $\delta K_{ij}$, what
makes a definition of quasilocal stress tensor more elusive. But one knows that
in AAdS spacetimes the leading order of
the asymptotic expansion in $K_{ij}$ coincides with the leading order of the induced metric $h_{ij}$ , i.e.,
\begin{equation}
K_{ij}=\frac{1}{\ell} \frac{g_{(0)ij}}{\rho}+{\cal O}(\rho).
\end{equation}
The above relation inspires a reformulation of holographic
renormalization in terms of an expansion of the extrinsic
curvature  \cite{Papadimitriou-Skenderis}. This suggests it might
be still possible to obtain a regularized stress tensor $\langle
T_{ij} \rangle$ associated to $g_{(0)ij}$  even though
Kounterterms regularization does not lend itself to a Brown-York
stress tensor definition
 $T_{ij}={2\over\sqrt{-h}}{\delta I_{reg}\over\delta h^{ij}}$. It also motivates a direct comparison
 with the standard procedure what, until now, has been performed in Einstein gravity only in three
 dimensions \cite{Miskovic-Olea}. For four and higher even dimensions, this is carried out below.

\section{Gauss-Bonnet invariant in 4-dimensional AdS gravity}

Let us consider the Einstein-Hilbert action with negative
cosmological
constant in four dimensions supplemented by the Gauss-Bonnet (GB) term
$\mathcal{E}_{4}$ with an arbitrary coupling constant $\alpha $%
\begin{equation}
I=\int\limits_{M}d^{4}x \sqrt{-{\cal G}}\left[\frac{1}{16\pi
G}\left( R-2\Lambda \right) +\alpha (R_{\mu \nu \alpha \beta }R^{\mu
\nu \alpha \beta }-4R_{\mu \nu }R^{\mu \nu }+R^{2})\right],
\label{I4}
\end{equation}%
where $\Lambda =-3/\ell ^{2}$ is the cosmological constant in terms
of the AdS radius $\ell $. It was shown in ref.\cite{ACOTZ4} that a
well-posed action principle for gravity with AdS asymptotics removes
the arbitrariness in the GB coupling. Since $\mathcal{E}_{4}$ is a
topological invariant, it does not modify the field equations.
However, it still contributes to the surface term when
the total action is varied
\begin{equation}
\delta I=\int\limits_{M}EOM+\int\limits_{\partial M}d^{3}x\,%
\sqrt{-h}\,n_{\sigma }\delta _{\lbrack \gamma \delta \alpha \beta
]}^{[\sigma \lambda \mu \nu ]}\mathcal{G}^{\delta \varepsilon
}\delta \Gamma _{\lambda \varepsilon }^{\gamma } \left(
\frac{1}{64\pi G}\delta _{\lbrack \mu \nu ]}^{[\alpha \beta
]}+\alpha R_{\mu \nu }^{\alpha \beta }\right) , \label{deltaI4}
\end{equation}%
where $n_{\mu }$ is the normal vector to the boundary \cite{delta}.  The total
action is rendered stationary demanding $\delta I=0$ on-shell for a given
boundary condition. For asymptotically locally AdS spacetimes, i.e.,
$R_{\mu \nu }^{\alpha \beta }+\frac{1}{\ell ^{2}}\delta _{\lbrack
\mu \nu ]}^{[\alpha \beta ]}=0$
at $\partial M$, the variational principle fixes the coupling constant as $%
\alpha =\ell ^{2}/(64\pi G)$, which produces finite Noether charges \cite{ACOTZ4}.
Surprisingly enough, the same value of $%
\alpha $ regularizes the Euclidean action in a background-independent way \cite%
{OleaJHEP} and casts eq.(\ref{I4}) into MacDowell-Mansouri form \cite%
{Mac-Man}%
\begin{equation}
I_{4}=\frac{\ell ^{2}}{256\pi G}\int\limits_{M}d^{4}x\,\sqrt{-%
\mathcal{G}}\,\delta _{\lbrack \gamma \delta \alpha \beta ]}^{[\sigma
\lambda
\mu \nu ]}\left( R_{\sigma \lambda }^{\gamma \delta }+\frac{1}{\ell ^{2}}%
\,\delta _{\lbrack \sigma \lambda ]}^{[\gamma \delta ]}\right)
\left( R_{\mu \nu }^{\alpha \beta }+\frac{1}{\ell
^{2}}\,\delta _{\lbrack \mu \nu ]}^{[\alpha \beta ]}\right) .
\label{MacMan}
\end{equation}%
 Using the
field equations, one proves that the Weyl tensor is
\begin{equation}
W_{\mu \nu }^{\alpha \beta }=R_{\mu \nu }^{\alpha \beta }+\frac{1}{\ell ^{2}}%
\,\delta _{\lbrack \mu \nu ]}^{[\alpha \beta ]},  \label{weyl4}
\end{equation}%
where the r.h.s. is the curvature of the AdS group (the rest
corresponds to the torsion, which vanishes in Riemann gravity).
This fact implies that the action (\ref{MacMan}) is on-shell
equivalent to conformal gravity
\begin{equation}
I_{4}=\frac{\ell ^{2}}{64\pi G}\int\limits_{ M}d^{4}x\,\sqrt{-%
\mathcal{G}}\,W_{\mu \nu \alpha \beta }W^{\mu \nu \alpha \beta },
\label{IregWeyl}
\end{equation}%
because any trace of $W_{\mu \nu \alpha \beta }$ is identically zero
\cite{MPT}.

In what follows, we show that the addition of a topological
invariant of the Euler class recovers the standard counterterm
regularization and holographic stress tensor, by considering its
equivalent boundary formulation.

\section{Boundary formulation }

In a four-dimensional manifold without boundaries, the integration
of the GB term is proportional to the Euler characteristic $\chi
(M)$. When a boundary is introduced, a correction to $\chi (M)$ is
required, such that the Euler theorem reads
\begin{equation}
\int\limits_{M}d^{4}x\,\mathcal{E}_{4}=32\pi ^{2}\chi
(M)+\int\limits_{\partial M}d^{3}x\,B_{3},  \label{EulerTh}
\end{equation}%
where $B_{3}$ is a boundary term known as Second Chern Form. If the
spacetime is foliated using Gaussian (radial) coordinates
$ds^{2}=N^{2}(\rho )d\rho ^{2}+h_{ij}(\rho ,x)dx^{i}dx^{j}$,
the term $B_{3}$ is given as a polynomial in the extrinsic curvature $K_{ij}=-\frac{%
1}{2N}\,\partial _{\rho }h_{ij}$ and the intrinsic curvature $\mathcal{R}%
_{kl}^{ij}(h)$ as \cite{OleaJHEP}%
\begin{equation}
B_{3}=4\sqrt{-h}\ \delta _{\left[ j_{1}j_{2}j_{3}\right] }^{\left[
i_{1}i_{2}i_{3}\right] }K_{i_{1}}^{j_{1}}\left( \frac{1}{2}\,\mathcal{R}%
_{i_{2}i_{3}}^{j_{2}j_{3}}(h)-\frac{1}{3}\,K_{i_{2}}^{j_{2}}K_{i_{3}}^{j_{3}}%
\right) .  \label{B3KR}
\end{equation}%
There is a reason why to consider the boundary formulation of
topological invariants beyond the purpose of comparison with the
counterterm regularization. The boundary dynamics does not tell
between the Euler and boundary term $B_{3}$, as they are locally
equivalent. However, computations of the Euclidean action show that
the Euler term shifts the black hole entropy $S$ by a constant
proportional to $\chi (M)$ \cite{OleaJHEP}, what can also be obtained using Wald's
entropy formula. Thus, $S$ may take negative values
for topological black holes with hyperbolic spatial section, what
can only be avoided by supplementing the action with the Kounterterm $B_{3}$ instead.

In order to compare to the standard regularization procedure,
one can simply add and subtract the Gibbons-Hawking term from the
Einstein-Hilbert action plus the boundary term $B_{3}$,
\begin{equation}
I_{4}=I_{EH}-\frac{1}{8\pi G}\int\limits_{\partial M}d^{3}x\sqrt{-h}%
\,K+\int\limits_{\partial M}d^{3}x\,\mathcal{L}_{ct}.
\label{addandsubGH}
\end{equation}%
The first two terms define the Dirichlet problem in gravity, while
the quantity $\mathcal{L}_{ct}$ is given by
\begin{equation}
\mathcal{L}_{ct}=\frac{\ell ^{2}}{16\pi G}\sqrt{-h}\delta _{\left[
j_{1}j_{2}j_{3}\right] }^{\left[ i_{1}i_{2}i_{3}\right]
}K_{i_{1}}^{j_{1}} \left( \frac{1}{2}\,\mathcal{R}_{i_{2}i_{3}}^{j_{2}j_{3}}(h)-\frac{1}{3}\,%
K_{i_{2}}^{j_{2}}K_{i_{3}}^{j_{3}}+\frac{1}{\ell ^{2}}\,\delta
_{i_{2}}^{j_{2}}\delta _{i_{3}}^{j_{3}}\right) . \label{Lct4K}
\end{equation}%
For the boundary metric $h_{ij}=g_{ij}/\rho$, the intrinsic curvature
and the determinant rescale as $\mathcal{R}%
_{kl}^{ij}(h)=\rho \mathcal{R}_{kl}^{ij}(g)$ and
$\sqrt{-h}=\sqrt{-g}/\rho ^{3/2}$, respectively. This also implies
\begin{equation}
K_{i}^{j}=K_{ik}h^{kj}= \frac{1}{\ell }\left(\delta _{i}^{j}-\rho
k_{i}^{j}\right)
\end{equation}%
for the extrinsic curvature, with the definition $k_{i}^{j}=g^{jk}\partial _{\rho }g_{ki}$.
 Expanding eq.(\ref{Lct4K}) in FG form, one notices that
$k_{j}^{i}$ is absent from the divergent terms,
\begin{equation}
\mathcal{L}_{ct}=\frac{1}{8\pi G}\frac{\sqrt{-g}}{\rho ^{3/2}}\left( \frac{2%
}{\ell }+\frac{\ell }{2}\,\rho \mathcal{R}(g)\right) +\mathcal{O}(\rho
^{1/2}) \label{Lct}
\end{equation}%
 such that one recovers the Balasubramanian-Kraus local counterterms%
\begin{equation}
\mathcal{L}_{ct}=\frac{1}{8\pi G}\sqrt{-h}\left( \frac{2}{\ell
}+\frac{\ell }{2}\mathcal{R}(h)\right) .  \label{LctBK}
\end{equation}%
The agreement with the standard holographic renormalization can be
also seen from the on-shell variation of the action (\ref{deltaI4}),
which for the radial foliation and the value $\alpha =\ell ^{2}/(64\pi G)$ adopts the form%
\begin{equation}
\delta I_{4}=\frac{\ell^{2}}{32\pi G}\int\limits_{\partial M}d^{3}x\sqrt{-h}%
\delta _{\left[ mnp\right] }^{\left[ jkl\right] } \left( \delta K_{j}^{m}+%
\frac{1}{2}K_{i}^{m}\left( h^{-1}\delta h\right) _{j}^{i}\right)
\left( R_{i_{2}i_{3}}^{j_{2}j_{3}}+\frac{1}{\ell ^{2}}\delta
_{\lbrack i_{2}i_{3}]}^{[j_{2}j_{3}]}\right) .  \label{varIT}
\end{equation}%
Expanding the fields in the FG frame, the first term vanishes at the
boundary, whereas the second gives a stress tensor%
\begin{equation}
\tau _{i}^{j}=\frac{\ell^{2}}{32\pi G}\,\delta _{\left[ mnp\right] }^{\left[ jkl%
\right] }K_{i}^{m}\left( R_{kl}^{np}+\frac{1}{\ell ^{2}}\delta
_{\lbrack kl]}^{[np]}\right). \label{stressK}
\end{equation}%
Note that any conserved quantity constructed with this stress tensor will
vanish for spacetimes which are globally of constant curvature, as
AdS vacuum. Using Gauss-Codazzi relations, one might also notice that
$\tau _{i}^{j}$ contains higher powers in the extrinsic curvature.
However, it is straightforward to prove that $\tau _{i}^{j}$
coincides up to the relevant order in $\rho $ with the
Balasubramanian-Kraus stress tensor $T_{i}^{j}$, when it is appropriately rewritten
\begin{eqnarray}
T_{i}^{j} &=& \frac{1}{8\pi G}\left( K_{i}^{j}-\delta _{i}^{j}K+\frac{2}{\ell }%
\delta _{i}^{j}-\left( \mathcal{R}_{i}^{j}(h)-\frac{1}{2}\delta _{i}^{j}%
\mathcal{R}(h)\right) \right) \nonumber \\
   &=& \frac{\rho \ell }{32\pi G}\,\delta _{\left[ inp\right] }^{\left[ jkl%
\right] }\left( \mathcal{R}_{kl}^{np}(g)+\frac{4}{\ell^{2}}\delta
_{k}^{n}k_{l}^{p}\right). \label{BKtensor}
\end{eqnarray}%

The above derivation shows that the divergence cancelation provided
by the counterterm series can be regarded as a {\it topological}
regularization, since it comes from the addition of the GB term with a coupling such that
the regularized action takes the MacDowell-Mansouri
form (\ref{MacMan}).

 In the
 holographic renormalization framework, the information on the
holographic stress tensor in four dimensions is carried by the
coefficient $g_{(3)}$ in FG expansion
\begin{equation}
 g_{ij}(x,\rho )=g_{(0)ij}+\rho g_{(1)ij}+\rho^{3/2} g_{(3)ij}+\cdots\,.
\end{equation}
It is just after solving the
Einstein equations order by order in $\rho$ that the vanishing of
the Weyl anomaly comes from the zero trace of $g_{(3)}$
\cite{dHSS}. On the other hand, the anomaly $\cal{A}$ can be also
read off from a Weyl transformation with infinitesimal parameter
$\sigma$ on the regularized action, that is,
$\delta_{\sigma}I_{reg}=\int_{\partial M} d^{D-1}x \sqrt{g_{(0)}}\,
\sigma\cal{A} $. This means that one might have concluded the
same by simple inspection of the eq.(\ref{IregWeyl}), since it is
manifestly invariant under conformal transformations.

Up to the relevant order, the stress tensor (\ref{stressK}) can be rewritten as
\begin{equation}
\tau _{i}^{j} =\frac{\ell }{8\pi G}W_{\,\mu i\nu }^{j}\,n^{\mu
}n^{\nu },
\end{equation}
using the traceless property and index symmetries of the Weyl
tensor.

 The conformal completion technique \cite%
{Ashtekar-Magnon-Das} defines an AAdS spacetime in such a way that the
metric ${\cal G}_{\mu \nu }$ which obeys the Einstein equations can
be conformally mapped into an \emph{unphysical} one ${\cal
\tilde{G}}_{\mu \nu}= \Omega^{-2}{\cal
{G}}_{\mu \nu}$ by a smooth conformal factor $\Omega$ which
satisfies precise fall-off conditions. The procedure gives rise to a
background-independent conserved charge for every asymptotic
symmetry $\xi ^{i}$ as the integral on the spatial section $\tilde{\Sigma}$ of the
boundary
\begin{equation}
{\cal H}_{\xi }=\frac{\ell }{8\pi G}\int\limits_{\tilde{\Sigma}}\tilde{E}%
_{i}^{j}\xi ^{i}\tilde{u}_{j}\,d\tilde{\Sigma}, \label{AMD}
\end{equation}
 where $\tilde{E}_{i}^{j}=\Omega ^{3-D}\tilde{W}_{\,\mu i\nu }^{j}\,%
\tilde{n}^{\mu }\tilde{n}^{\nu }/(D-3)$  is the {\it electric} part
of the unphysical Weyl tensor, $d\tilde{\Sigma}$ is the integration
element on $\tilde{\Sigma}$  and $\tilde{u}_{j}$ is the unit
timelike normal to $\tilde{\Sigma}$. Rescaling all the quantities
into the ones of the spacetime metric,  it is easy to prove that the
conserved quantities $Q_{\xi }\equiv\int_{\Sigma }\tau
_{i}^{j}\xi ^{i}u_{j}\,d\Sigma$ coming from Kounterterms
regularization in $D=4$ are the same as the Ashtekar-Magnon-Das
 formula (\ref{AMD}).

\section{Pontryagin term and self-dual solutions}

In four dimensions there exists an additional (odd parity)
topological invariant known as Pontryagin term $\mathcal{P}_{4}$,
which is locally equivalent to the derivative
of the gravitational Chern-Simons term
\begin{equation}
\mathcal{P}_{4}=-\frac{1}{4}\epsilon ^{\mu \nu \alpha \beta }R_{\mu
\nu }^{\sigma \lambda }R_{\sigma \lambda \alpha \beta }=
\epsilon^{\mu \nu \alpha \beta }\partial_{\mu}\left( \Gamma
_{\nu\lambda }^{\sigma}\partial_{\alpha}\Gamma _{\beta\sigma
}^{\lambda }+\frac{2}{3}\Gamma _{\nu\lambda }^{\sigma}\Gamma
_{\alpha\varepsilon }^{\lambda }\Gamma _{\beta \sigma }^{\varepsilon
}\right), \label{pontryagin}
\end{equation}%
where $\epsilon ^{\mu \nu \alpha \beta }$ is the constant
Levi-Civita tensor density.

 The Pontryagin term $F\wedge F$ in
four-dimensional Maxwell electromagnetism modifies the dynamics such
that the Lorentz boost and the parity invariance are lost when it is
coupled through an
external, fixed quantity.

We will consider here the addition of the Pontryagin term with a
constant coupling  $\beta$ to the regularized action, i.e.,
\begin{equation}
\tilde{I}=I_{EH}+\frac{\ell^2}{64\pi G} \int\limits_{M}
d^4x\,{\cal{E}}_{4}+\beta \int\limits_{M} d^4x\,{\cal{P}}_{4}\,,
\label{Itilde}
\end{equation}
with Euclidean signature.
 Therefore, in a similar fashion to the
case of the addition of the Euler term, the bulk dynamics cannot fix
the Pontryagin coupling. However, one may expect that again the
variational principle would provide a criterion to remove the
arbitrariness
in $\beta$.

The on-shell variation of the total action produces %
\begin{equation}
\delta \tilde{I} =\int\limits_{\partial
M}d^{3}x\,\sqrt{\cal{G}}\,\frac{n_{\sigma }}{N}\,\delta
\Gamma _{\varepsilon \lambda }^{\gamma }\left( \frac{\ell ^{2}}{64\pi G}\,%
\delta _{\lbrack \gamma \delta \alpha \beta ]}^{[\sigma \lambda \mu \nu ]}%
{\cal G}^{\delta \varepsilon }W_{\mu \nu }^{\alpha \beta } + \beta
\,\frac{\epsilon ^{\sigma \lambda \mu \nu }}{\sqrt{\cal{G}}}\,{\cal
G}_{\gamma \tau }W_{\mu \nu }^{\varepsilon \tau }\right) ,
\label{varItilde}
\end{equation}%
where in the last term the part along $\delta_{[\mu \nu]
}^{[\varepsilon \tau] }$ is identically zero. The total surface
term must vanish identically for certain boundary conditions. The
argument here is different from the one used to fix the GB
coupling in eq.(\ref{I4}). In that case, $\alpha $ is also determined
 from the cancellation of the leading-order divergences in
the Euclidean action what can be seen, e.g., from evaluating it for
Schwarzschild-AdS black hole
\begin{equation}
-TI_{SAdS}=TS-\frac{\pi
r^3}{4G\ell^2}\left(1-\frac{64\pi G}{\ell^2}\,\alpha\right)
-\frac{M}{2}\left(1+\frac{64\pi G}{\ell^2}\,\alpha\right),
\label{IESAdS}
\end{equation}
where $S$ and $M$ are the black hole entropy and mass, respectively, and $T$ is the Hawking temperature.
It is clear that the correct black hole
thermodynamics is reproduced only by the same value of $\alpha$ as before.
Moreover, for a given cosmological constant, it is not possible to express the variation
eq.(\ref{varItilde}) only in terms of the Weyl tensor unless $\alpha$ takes
 the value fixed in the previous sections.

The result (\ref{IESAdS}) remains unchanged when ${\cal{P}}_{4}$
is added to the action, as it vanishes
for static AdS$_{4}$ black holes.
In general, it can be shown that the contribution of the Pontryagin term to
the action is at most finite.

This means that we should look
for asymptotic conditions in the next-to-leading order in the
curvature of the AdS group (\ref{weyl4}).
 Considering (anti) self-duality in the Weyl tensor
\begin{equation}
W_{\mu \nu \alpha \beta }=\pm \frac{1}{2}\sqrt{\cal{G}}\,\epsilon
_{\mu \nu \lambda \sigma }W_{\alpha \beta }^{\lambda \sigma }
\label{selfdualC}
\end{equation}
in the asymptotic region, we can fix the coupling constant of $\mathcal{P}_{4}$ as
\begin{equation}
 \beta =\pm \frac{\ell ^{2}}{32\pi G}\,,
\label{betadual}
\end{equation}
demanding a well-posed action principle.

For arbitrary $\beta$, the variation of the action (\ref{Itilde})
projected to the boundary indices defines a total stress tensor ${\cal T}_{j}^{i}$
\begin{eqnarray}
\delta \tilde{I}&=&
\frac{1}{2}\int\limits_{\partial
M}d^{3}x\,\sqrt{h}\,{\cal T}_{j}^{i} (h^{-1}\delta h)_{i}^{j}
\nonumber \\
&=& \frac{1}{2}\int\limits_{\partial M}d^{3}x\,\sqrt{h}
\left( T_{j}^{i}+\beta \, C_{j}^{i}\right) (h^{-1}\delta h)_{i}^{j}\,,
\label{totalstress}
\end{eqnarray}
where $T_{j}^{i}$ is the stress tensor (\ref{BKtensor}) and
$C^{i}_{j}=\frac{1}{\sqrt{h}}\,\epsilon^{ikl}\nabla _{k}\left( \mathcal{R}_{lj}-\frac{%
1}{4}h_{lj}\mathcal{R}\right) $ is the Cotton tensor, obtained from
the functional variation of the gravitational Chern-Simons term
respect to the induced metric $h_{ij}$. The Cotton tensor is
symmetric, traceless and covariantly conserved, and contributes as
above to the total stress tensor of the theory when $h_{ij}$ is held
fixed on the boundary (Dirichlet problem).

The term ${\cal{P}}_{4}$ does not modify the AdS asymptotics, such that we can
use FG expansion and find the finite part of eq.(\ref{totalstress}), which is given by
\begin{equation}
\delta \tilde{I}=\frac{1}{2}\int\limits_{\partial M}d^{3}x\,\sqrt{g_{(0)}}
\left( -\frac{3}{16\pi G\ell }\,g_{(3)i}^{j}+\beta \, C_{j}^{i}(g_{(0)})\right)
(g_{(0)}^{-1}\delta g_{(0)})_{i}^{j}\,.
\label{totalholstress}
\end{equation}

In a similar fashion, (anti) self-duality reads
 \begin{eqnarray}
&&\rho\, {\cal W}_{kl}^{np}+\frac{3\rho ^{3/2}}{2\ell ^{2}}\left(
g_{(3)k}^{n}\delta _{l}^{p}-g_{(3)l}^{n}\delta _{k}^{p}+\delta
_{k}^{n}g_{(3)l}^{p}-\delta _{l}^{n}g_{(3)k}^{p}\right)+\mathcal{O}(\rho ^{2}) \nonumber \\
&&\quad\qquad=\mp \,\frac{\rho ^{3/2}}{\ell \sqrt{g_{(0)}}}\,\epsilon ^{npm}
\left(\nabla _{(0)k}g_{(1)ml}-\nabla _{(0)l}g_{(1)mk}\right)+\mathcal{O}(\rho ^{2})\,,
\end{eqnarray}
where ${\cal W}$ is the Weyl tensor of $g_{(0)}$.

As a consequence, when the condition (\ref{selfdualC}) holds,
the value $\beta=\pm\ell^2/(32\pi G)$ corresponds to the self-dual point
where the total stress tensor vanishes identically, i.e., ${\cal T}_{j}^{i}=0$.

This reproduces the relation between the holographic stress tensor $T_{j}^{i}$
and the Cotton tensor
\begin{equation}
T_{j}^{i}=\pm\frac{\ell ^{2}}{8\pi G}\,C_{j}^{i}\,,  \label{TCduality}
\end{equation}
which has been observed in recent literature for solitonic solutions \cite{deHaro-Petkou},
electric-magnetic transformations in the fields in first-order gravity \cite{MPT}
and axial-polar perturbations in hydrodynamic models in AdS$_{4}$ \cite{Bakas}.

The full duality between the renormalized stress tensor and Cotton tensor has been obtained
in \cite{deHaro} by relating two dual boundary CFTs which correspond to Dirichlet
and Neumann boundary conditions (for a related work on boundary conditions, see, \cite{Compere-Marolf}).
The two descriptions are mapped one into another by a Legendre
transformation generated by a gravitational Chern-Simons term.

The total action for the particular value of $\beta$ which realizes
the relation (\ref{TCduality}) can be written in tetrad formalism as
\begin{equation}
\tilde{I}=\frac{\ell ^{2}}{64\pi G}\int\limits_{M}\left( \epsilon
_{ABCD}W^{AB}W^{CD}\mp 2W^{AB}W_{AB} \right), \label{Iforms}
\end{equation}
with the Weyl 2-form $W^{AB}=W^{\alpha \beta}_{\mu \nu}
e^{A}_{\alpha} e^{B}_{\beta} dx^{\mu}dx^{\nu}$ in terms of the local
orthonormal basis $e^{A}=e^{A}_{\mu}dx^{\mu}$.  In this notation the
(anti) self-duality condition (\ref{selfdualC}) reads
$W_{AB}=\pm \ast W_{AB}=\pm \frac{1}{2}\epsilon _{ABCD}W^{CD}$, with
 $\ast \ast =+1$ for Euclidean signature.

Using the identity $\epsilon_{ABCD}W^{AB}W^{CD}=\frac{1}{2}\epsilon
_{ABCD}\left( W^{AB}W^{CD}+\ast W^{AB}\ast W^{CD}\right)$ and also
that $\epsilon _{ABCD}W^{AB}\ast W^{CD}=-2{\cal{P}}_{4}d^4x$ (in an
analogous way as in Yang-Mills theory), the total action
(\ref{Iforms}) can be cast into the form
\begin{eqnarray}
\tilde{I}&=&\frac{\ell ^{2}}{128\pi G}\int\limits_{M}\epsilon _{ABCD}\left(
W^{AB}\mp \ast W^{AB}\right) \left( W^{CD}\mp \ast
W^{CD}\right) \nonumber\\
 &=& \frac{\ell ^{2}}{16\pi G} \int\limits_{M} \sqrt{\det \left(W^{AB}\mp \ast W^{AB}\right)}.
\label{dual-anti-action}
\end{eqnarray}%

It is evident from the form of eq.(\ref{MacMan})
that the value of the action reaches an absolute minimum for spacetimes which are
globally of constant curvature (vacuum states of AdS gravity). The action (\ref{dual-anti-action})
naturally generalizes this property to states which are globally
(anti) self-dual in AdS gravity.

\section{Conclusions}
We have shown that the standard regularization of AdS gravity with
counterterms is indeed topological, as it can be obtained from the
addition of the Gauss-Bonnet invariant or the corresponding boundary
term.

 We have also considered the odd parity Pontryagin invariant,
which accounts for viscosity in hydrodynamic models and for
\emph{magnetic} properties of solitonic solutions in AdS$_{4}$ (by
analogy to the charge formula (\ref{AMD}) which involves the
electric part of the Weyl tensor due to the addition of the Euler
term). It is shown that the inclusion of this term is consistent
assuming an asymptotic (anti) self-dual condition on the Weyl
tensor. This reasoning explains the holographic stress
tensor/Cotton tensor relation (\ref{TCduality}) recently found in different setups in
the literature, and interprets it as coming from a duality between
topological invariants.

 The addition of topological invariants of the
Euler class to the Einstein-Hilbert gravity action in $D=2n$
dimensions  was studied in ref.\cite{ACOTZ2n}, with the purpose of rendering
finite the Noether charges for AAdS spacetimes. The variational
principle singles out the value of the Euler coupling which produces
a regularizing effect. One can instead consider the action
supplemented by a boundary term
$I_{2n}=I_{EH}+c_{2n-1}\int_{\partial M}d^{2n-1}x B_{2n-1}$,
where $c_{2n-1}$ is a constant. In ref.\cite{OleaJHEP}, it is
claimed that the term $B_{2n-1}$ which solves the regularization
problem in even-dimensional AdS gravity is always prescribed by the
Euler theorem and written using a parametric integration as a
polynomial in the intrinsic and extrinsic curvatures
\begin{eqnarray}
B_{2n-1} &=&2n\sqrt{-h}\int\limits_{0}^{1}dt\ \delta _{\left[
j_{1}\cdots j_{2n-1}\right] }^{\left[ i_{1}\cdots
i_{2n-1}\right] }K_{i_{1}}^{j_{1}}\left( \frac{1}{2}\mathcal{R}%
_{i_{2}i_{3}}^{j_{2}j_{3}}-t^{2}K_{i_{2}}^{j_{2}}K_{i_{3}}^{j_{3}}\right)
\times \cdots   \notag \\
&&\cdots \times \left( \frac{1}{2}\mathcal{R}%
_{i_{2n-2}i_{2n-1}}^{j_{2n-2}j_{2n-1}}-t^{2}K_{i_{2n-2}}^{j_{2n-2}}K_{i_{2n-1}}^{j_{2n-1}}\right)
\label{B2n-1}
\end{eqnarray}%
with a coupling constant $c_{2n-1}=(-\ell
^{2})^{n-1}/(16\pi Gn(2n-2)!)$. On purpose, we have not absorbed the
constant in the boundary term, in order to stress the geometrical
origin of the Kounterterm $B_{2n-1}$, as it is linked to topological invariants.

Furthermore, it has been proved that the term (\ref{B2n-1})
regulates the Euclidean action and conserved quantities in any
gravity theory of Lovelock type with AdS asymptotics -- including
Einstein-Gauss-Bonnet AdS--, and where the information on a
particular theory is contained only in its coupling constant
\cite{EGB-KO}.

As in the four-dimensional case, we add
and subtract the Gibbons Hawking term, i.e., $%
I_{2n}=I_{EH}-\frac{1}{8\pi G}\int_{\partial M}d^{2n-1}x\sqrt{-h}%
K+\int_{\partial M}d^{2n-1}x\mathcal{L}_{ct}$, where
$\mathcal{L}_{ct}$ is
given by%
\begin{eqnarray}
\mathcal{L}_{ct} &=&\frac{(-\ell ^{2})^{n}}{8\pi G(2n-2)!}\sqrt{-h}\delta _{%
\left[ j_{1} \cdots j_{2n-1}\right] }^{\left[ i_{1}\cdots i_{2n-1}
\right] }K_{i_{1}}^{j_{1}} \int\limits_{0}^{1}dt\ \left[ \left( \frac{1}{2}\mathcal{R}%
_{i_{2}i_{3}}^{j_{2}j_{3}}-t^{2}K_{i_{2}}^{j_{2}}K_{i_{3}}^{j_{3}}\right)
\cdots \right. \notag
\\
&& \quad \cdots \left. \left( \frac{1}{2}\mathcal{R}%
_{i_{2n-2}i_{2n-1}}^{j_{2n-2}j_{2n-1}}-t^{2}K_{i_{2n-2}}^{j_{2n-2}}K_{i_{2n-1}}^{j_{2n-1}}\right) +%
\frac{(-1)^{n}}{\ell ^{2n-2}}\delta _{i_{2}}^{j_{2}}\cdots \delta
_{i_{2n-1}}^{j_{2n-1}}\right] .
\end{eqnarray}%
When all the fields are expanded in FG frame, one can collect terms
as a power series in $\rho $ and perform explicitly the parametric
integration. It is useful to express the extrinsic curvature
expansion as $K_{j}^{i}=\frac{1}{\ell }\delta _{j}^{i}-\rho \ell S_{j}^{i}(g)+\mathcal{O}%
(\rho ^{2})$, where $S_{j}^{i}(g)=\frac{1}{D-3}(\mathcal{R}_{j}^{i}(g)-\frac{1}{2(D-2)}%
\delta _{j}^{i}\mathcal{R}(g))$ is the Schouten tensor of the metric $%
g_{ij}(x,\rho )$. Owing to the rescaling properties of the boundary
Riemann tensor, the result can be written as a series of intrinsic
counterterms

\begin{eqnarray}
\mathcal{L}_{ct} &=&\frac{\sqrt{-h}}{8\pi G}\left[ \frac{(2n-2)}{\ell }+%
\frac{\ell }{2(2n-3)}\,\mathcal{R}+\right. \nonumber  \\
&&\hspace{-5mm}\left. +\frac{\ell ^{3}}{2(2n-3)^{2}(2n-5)}\left( 2\mathcal{R}^{ij}%
\mathcal{R}_{ij}-\frac{(2n+1)}{4(2n-2)}\mathcal{R}^{2}-\frac{(2n-3)}{4}%
\mathcal{R}^{ijkl}\mathcal{R}_{ijkl}\right) +\cdots \right] .
\end{eqnarray}%
which includes a rather unusual $(Riemann)^{2}$ contribution.
However, the fall-off conditions for AAdS solutions imply that the  Weyl tensor is such that $\sqrt{-h}%
W^{ijkl}W_{ijkl}\sim \frac{1}{r^{D-1}}$ in Schwarzschild-like
coordinates (see also \cite{Ashtekar-Magnon-Das}). Using this
property for $D\geq6$, we trade off the Riemman-squared term for the
other curvature-squared
terms, i.e., $\mathcal{R}^{ijkl}\mathcal{R}_{ijkl}=\frac{4}{(2n-3)}(\mathcal{%
R}^{ij}\mathcal{R}_{ij}-\frac{1}{2(2n-2)}\mathcal{R}^{2})$.
Remarkably enough, the series $\mathcal{L}_{ct}$ adopts the form of
standard counterterms
obtained by holographic renormalization%
\begin{eqnarray}
\mathcal{L}_{ct} &=&\frac{\sqrt{-h}}{8\pi G}\left[ \frac{(2n-2)}{\ell }+%
\frac{\ell }{2(2n-3)}\mathcal{R}+\right. \nonumber \\
&&\left. +\frac{\ell ^{3}}{2(2n-3)^{2}(2n-5)}\left( \mathcal{R}^{ij}\mathcal{%
R}_{ij}-\frac{(2n-1)}{4(2n-2)}\mathcal{R}^{2}\right) +\cdots \right]
,
\end{eqnarray}%
where cubic terms in the curvature are required by the
regularization problem only for $D\geq 8$ dimensions.

 We will provide the details of this derivation in a forthcoming publication.

\section*{Acknowledgments}

We would like to thank A. Anabal\'on, D. Klemm, M. Leoni and S.
Theisen for interesting discussions and S. de Haro for useful
correspondence. R.O. also thanks Prof. S. Theisen for kind
hospitality at AEI, Golm, during the completion of this work. O.M.
is supported by FONDECYT grant 11070146 and the PUCV through the
project 123.797/2007. The work of R.O. was funded in part by
AEI-MPG.



\begin{thebibliography}{99}
\bibitem{delta} Totally antisymmetric delta symbol
is defined as the determinant of Kronecker deltas $\delta^{\mu}_{\nu}$ with normalization equal to unity.
\end{thebibliography}
\end{document}